\documentclass[superscriptaddress,aps,prl,twocolumn,showpacs]{revtex4}
\usepackage{times}
\usepackage{graphicx}

\input epsf

\begin{document}

\title{An accelerator mode based technique for studying quantum chaos}
\author{M.B.\thinspace d'Arcy}
\affiliation{Clarendon Laboratory, Department of
Physics,University of Oxford, Parks Road, Oxford, OX1 3PU, United
Kingdom}
\author{R.M.\thinspace Godun}
\affiliation{Clarendon Laboratory, Department of
Physics,University of Oxford, Parks Road, Oxford, OX1 3PU, United
Kingdom}
\author{D.\thinspace Cassettari}
\affiliation{Clarendon Laboratory, Department of
Physics,University of Oxford, Parks Road, Oxford, OX1 3PU, United
Kingdom}
\author{G.S.\thinspace Summy}
\affiliation{Clarendon Laboratory, Department of
Physics,University of Oxford, Parks Road, Oxford, OX1 3PU, United
Kingdom} \affiliation{Department of Physics, Oklahoma State
University, Stillwater, Oklahoma 74078-3072}
\date{\today}

\begin{abstract}
We experimentally demonstrate a method for selecting small regions
of phase space for kicked rotor quantum chaos experiments with
cold atoms. Our technique uses quantum accelerator modes to
selectively accelerate atomic wavepackets with localized spatial
and momentum distributions. The potential used to create the
accelerator mode and subsequently realize the kicked rotor system
is formed by a set of off-resonant standing wave light pulses. We
also propose a method for testing whether a selected region of
phase space exhibits chaotic or regular behavior using a Ramsey
type separated field experiment.
\end{abstract}


\maketitle

Chaos in quantum mechanics is still a relatively poorly defined
concept. Typically it is taken to refer to the behavior of a
quantum system which in the classical limit exhibits an
exponential sensitivity to initial conditions. A much studied
example of such a system is the delta-kicked rotor
\cite{Lichtenberg and Lieberman,Casati}. A realization of its
quantum version, in the guise of cold atoms exposed to the
ac-Stark shift potential of an off-resonant standing wave of light
\cite{Raizen 1st QDKR}, has elucidated many of the concepts
associated with quantum chaos \cite{ResonancePRL,Tunnelling2,Quantum-Classical,Multi-dimensions}%
. However, one substantial problem remains:\ there is no clear way
of distinguishing regular from chaotic dynamics. In the quantum
regime it is not possible to define a chaotic region of phase
space as being one in which two initially similar states have an
overlap which decreases exponentially with time. The unitary
nature of any interaction necessarily implies that the overlap
between such states remains unchanged. It has been suggested by
Peres \cite{Peres} that an alternative definition is needed for
quantum systems. Peres' proposal involves examining the evolution
of a state which can interact with two potentials which differ
very slightly in form. If the overlap between states produced by
an evolution under each potential decreases exponentially as a
function of time, then the region of phase space occupied by the
initial state can be said to be chaotic in the quantum sense. The
justification for this definition is that perturbation theory only
converges when the region under examination has regular properties
\cite{Feingold}. More recently Gardiner \textit{et al.}
\cite{Gardiner97} have proposed an experiment with a trapped ion
which would implement the essential features of Peres' idea.

In this paper we address two problems related to the experimental
determination of a phase space stability map for the atom optical
version of the QDKR. Firstly, we demonstrate a method for
preparing cesium atoms in a restricted region of phase space using
the same standing wave light pulses which create the rotor.
Secondly, we show how by examining the overlap between the
wavefunctions of the two hyperfine levels of the $6^{2}S_{1/2}$
cesium ground state it should be possible to test the type of
dynamics exhibited by the prepared atoms.

The basis of our atom optical version of the QDKR is exposure of laser
cooled cesium atoms to pulses (duration $t_{{\rm p}}$ and separation time $T$%
) of a standing wave of off-resonant light. The pulses are short
enough to allow the effect of the atoms's kinetic energy to be
neglected during a pulse. This places our experiment in the
Raman-Nath regime in which the spatially periodic ac-Stark shift
potential created by the light acts as a thin phase grating for
the atoms \cite{AccModePRL}. Thus an incident plane de Broglie
wave is diffracted into a series of \ ``orders'' separated in
momentum by $\hbar G$, where $G=2k_{{\rm L}}$ ($k_{{\rm L}}$ is
the light wavevector). One of the striking features of this system
is the existence of quantum resonances \cite
{ResonancePRL,newfishman}. These resonances occur when the pulse
interval is a multiple of the half-Talbot time, $T_{\frac12}=2\pi
m/\hbar G^{2}$. During these special times all of the diffraction
orders formed from incident plane waves with certain momenta will
freely evolve phases which are multiples of $2\pi $. For cesium,
the first quantum resonance occurs at $T=67$ $\mu $s.

\begin{figure}[tbp]
\begin{center}\mbox{ \epsfxsize 3.0in\epsfbox{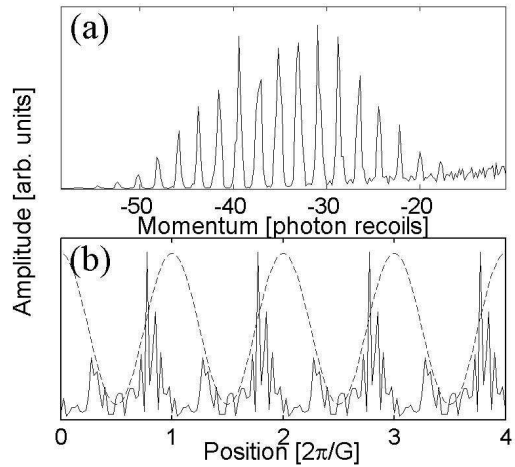}}\end{center}
\caption{Numerically simulated momentum (a) and spatial (b)
distributions of the accelerated component of an ensemble of cold
cesium atoms after 20 pulses of an off-resonant standing light
wave. The pulses had a separation of 60 $\protect\mu $s. The
momentum distribution consists of a comb of peaks separated by a
little more than two photon recoils. The degree of spatial
localization is consistent with an accelerator mode which contains
approximately three diffraction orders. The dashed line in (b)
shows the position of the standing light wave.}
\label{theorydistribution}
\end{figure}

In addition to being a paradigm of experimental quantum chaos,
this system can also produce quantum accelerator modes
\cite{AccModePRL}. This is achieved by adding a potential of the
form $U_{{\rm A}}(x)=max$, where $m$ is the atomic mass, $a$ is an
applied acceleration and $x$ is the position along the standing
wave. In our experiment the standing wave is oriented in the
vertical direction, so $a$ is the acceleration due to gravity.
Quantum accelerator modes are characterized by a fixed momentum
transfer of (on average) $\xi \hbar G$ during each standing wave
pulse (see Fig.\thinspace 1 of Ref.\thinspace \cite{AccModePRL}).
Typically the accelerator modes are formed when the value of $T$\
is near to a quantum resonance. For simplicity, we will henceforth
confine our attention to pulse repetition times near to the first
quantum resonance. One way of modelling the accelerator modes is
to make the approximation that they consist of just a few
diffraction orders (say $q-1$, $q$ and $q+1$, where $q$ is an
integer) which in the time between two light pulses accumulate a
phase difference which is very close to an integer multiple of
$2\pi $. At the next pulse this makes it possible for interference
to occur between the diffraction orders in such a way that the
population of the three orders centered on $q+\xi $
\cite{non-integer am} is enhanced. By using this rephasing
condition, it can be shown \cite{AccModePRA}\ that after $N_{{\rm
p}}$ pulses the central momentum of the accelerator mode is
given by $q=\frac{N_{{\rm p}}}{\gamma }\frac{\alpha ^{2}}{1-\alpha }=N_{{\rm %
p}}\xi $, where $\gamma =\hbar ^{2}G^{3}/2\pi m^{2}a$ and $\alpha
=T/T_{\frac12}$. The same rephasing condition also leads to the
conclusion that only incident plane waves with momenta
\begin{equation}
p_{{\rm init}}=\left( \alpha^{-1} l + {\rm const} \right) \hbar G
\label{comb eqn}
\end{equation}
(where $l$ is any integer) can ever participate in an accelerator
mode. Thus in momentum space an accelerator mode resembles a comb
with a tooth spacing of $\alpha^{-1} \hbar G$. Furthermore, since
plane waves separated by this momentum behave identically under
the action of the kicks, the accelerator mode effectively contains
only a single momentum. Figure \ref{theorydistribution}(a) shows
the theoretical momentum distribution of the accelerator mode
after one set of pulses, as calculated using a model based on
diffraction \cite{AccModePRL}. To reflect the experimental
situation the starting distribution was gaussian with a width of
12 $\hbar k_{\rm{L}}$ at FWHM. Although there is good qualitative
agreement between Eq.\thinspace (\ref{comb eqn}) and the
numerically derived momentum distribution, an explanation for the
finite widths of the comb elements is needed. To understand this
effect recall that each comb element must give rise to a set of
diffraction orders which are always spaced by exact multiples of
$\hbar G$. To determine what the real momentum distribution looks
like we must add together the diffraction from all the different
comb elements. These have a spacing of $\alpha^{-1} \hbar G$,
which for pulse interval times just less than the Talbot time is
slightly greater than $\hbar G$. The most obvious effect of
including the diffraction orders is that the width of each comb
element of the resultant distribution becomes non-zero and
increases with $|\alpha-1|$. Additionally, when all the
diffraction orders are weighted by the intensity of the comb
element from which they originated, the nearest-neighbor spacing
of the peaks in the momentum distribution is reduced as one moves
away from the center of the distribution. The opposite effect is
produced when the pulse interval is slightly greater than the
Talbot time. Although the change in spacing is not immediately
obvious in Fig.\thinspace \ref{theorydistribution}(a), it has been
confirmed in a detailed analysis of this data.

In Fig.\thinspace \ref{theorydistribution}(b) we show the spatial
form of the accelerator mode wavefunction as calculated with our
numerical model. Such a distribution is similar to that deduced
from the assumption that the mode consists of the sum of three
plane waves. Importantly, the wavefunction is periodically
localized in position with maxima occurring every $\lambda_{\rm
spat}=2\pi /G$. Since this is the wavelength of the standing wave
potential, points having this separation behave equivalently and
the wavefunction effectively has a {\em spread} in position which
is less than $\lambda_{\rm spat}$. From Fig.\thinspace
\ref{theorydistribution} we expect the width of one element of the
momentum comb to be approximately 0.4 $\hbar G$ and the extent of
each region of strong spatial localization to be $\lambda_{\rm
spat}/3$. For comparison, the extent in momentum of a unit cell of
the classical phase space is $(T_{1/2}/T) \hbar G$ while that in
position is $\lambda_{\rm spat}$. Hence the accelerator mode
isolates a restricted region in phase space which is about 10\% of
the overall area. Additionally, since the effect of the
accelerator mode is to produce a large momentum offset, it should
be straightforward to isolate these atoms.

We now discuss the experimental observation of the distributions
described above. Our atomic source was a cesium magneto-optic trap
containing approximately 10$^{7}$ atoms. After the trap was
switched off the atoms were cooled in an optical molasses to a
temperature of 5 $\mu $K, corresponding to a momentum width of 12
$\hbar k_{{\rm L}}$ at FWHM. Following release, the atoms were
exposed to a series of pulses from a vertically oriented standing
light wave, detuned $30$ GHz below the $F=4\rightarrow F^{\prime
}=3$ line
in the D1 transition at 894 nm. The pulses had a duration of $t_{{\rm p}%
}=0.5\;\mu $s and a peak intensity of\ approximately 20
W/cm$^{2}$. The atoms then fell freely until they passed through a
probe laser beam resonant with the D2 cycling transition. By
measuring the absorption of the probe light as a function of time
we were able to determine the momentum distribution of the atoms
in the $F=4$ level. Additional details of our apparatus can be
found in Ref.\thinspace \cite{AccModePRA}.

The momentum resolution of our experiment was approximately 2
photon recoils, determined by the initial spatial extent of the
optical molasses and the thickness of the probe laser beam. Hence
direct observation of the accelerator mode momentum comb was not
possible. Instead we used two sets of temporally separated light
pulses to infer its existence. The first set contained 20 pulses
and was used to create an accelerator mode. The resultant atomic
distribution was then translated in momentum by allowing the atoms
to fall for a variable amount of time $t_{{\rm wait}}$. Finally, a
second set of pulses, identical to the first, was applied. Figure\
\ref{momentum fringes} shows the resultant momentum distributions
as a function of $t_{{\rm wait}}$. Both panels contain a large
fraction of unaccelerated atoms near $p=0$ (dotted line), a group
of atoms that have been accelerated by one set of pulses
(dot-dashed line), and atoms that have been accelerated by both
pulse sets (dashed
line). In Fig.\thinspace \ref{momentum fringes}(a) the pulses had an interval of 60 $%
\mu $s and the accelerator mode imparted momentum in the negative direction
(with the convention gravity is negative), while in Fig.\thinspace \ref
{momentum fringes}(b) the interval was 74 $\mu $s and the accelerator mode
imparted positive momentum. The effect of $t_{{\rm %
wait}}$ is to allow gravity to translate the distribution in
momentum space. If any of this distribution overlaps with the
momentum comb of Eq.\thinspace (\ref{comb eqn}) when the second
set of pulses occurs then a further acceleration takes place. The
periodic variation in the doubly accelerated population with
$t_{\rm wait}$ is just the length of time required for gravity to
accelerate the atoms by the momentum separation of adjacent comb
elements, $\alpha ^{-1}\hbar G$. In the T=60 $\mu$s case this
gives a theoretical value of 753 $\mu$s between accelerator mode
revivals. Along the dashed line of Fig.\thinspace \ref{momentum
fringes}(a) we observe 763 $\pm$ 4 $\mu$s, the discrepancy with
the calculated value most likely due to the slight variation in
comb spacing across the accelerator mode discussed previously. A
similar level of agreement is found for T=74 $\mu$s. As regards
the width of each momentum comb element, we note that numerical
simulations have provided almost identical results to those shown
in Fig.\thinspace \ref{momentum fringes} so it is reasonable to
assume that the actual extent of each momentum element is as shown
in Fig.\thinspace \ref{theorydistribution}(a).

\begin{figure}[tbp]
\begin{center}\mbox{ \epsfxsize 3.0in\epsfbox{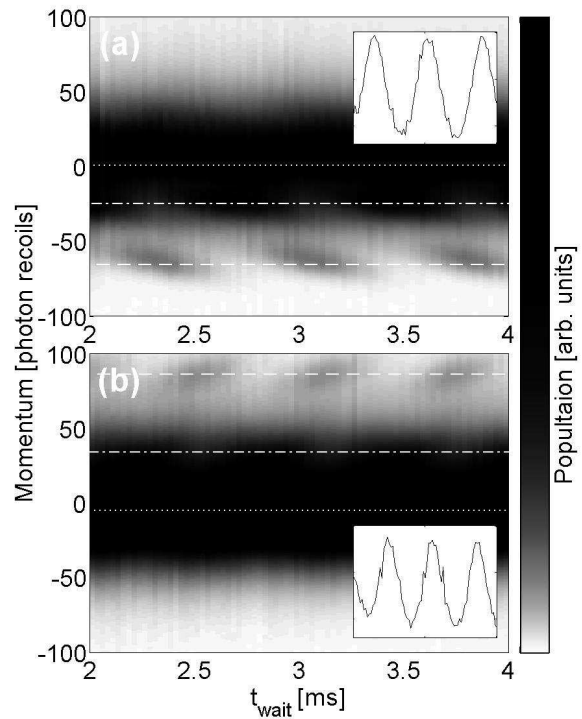}}\end{center}
\caption{Experimental momentum distributions of an ensemble of
cold cesium atoms after two sets of 20 pulses from an off-resonant
standing light wave. There was a variable amount of time $t_{{\rm
wait}}$ between the pulse sets. The dotted, dot-dashed and dashed
lines denote atoms accelerated by zero, one and two sets of pulses
respectively. The pulse spacing was $T=60$ $\mu $s in (a) and
$T=74$ $\mu $s in (b). Each vertical slice is the momentum
distribution obtained for one value of $t_{{\rm wait}}$ where the
degree of shading indicates the population. Gravity is in the
negative direction. The insets plot the data along the dashed
lines (atoms accelerated by both pulses).} \label{momentum
fringes}
\end{figure}

Although we have accounted for the accelerator mode revivals, we
have said nothing about their sloping orientation (as seen in
Fig.\thinspace \ref{momentum fringes}), nor about this slope's
dependence on pulse interval. An explanation for this effect can
be found by returning to the earlier observation that the momentum
intervals between the peaks of Fig.\thinspace
\ref{theorydistribution}(a) change as one moves away from the
center of the accelerator mode. This implies that not all parts of
the distribution have to be translated by the same amount to
regain the accelerator mode condition of Eq.\thinspace (\ref{comb
eqn}). At 60 $\mu $s, the outer peaks bulge towards the center of
the distribution. Assuming that the whole distribution is
accelerating under gravity, the first part to regain the
accelerator mode condition will be the component with the greatest
positive momentum. For 74 $\mu $s the opposite will be the case
and the direction of the revival's slope will flip.

\begin{figure}[tbp]
\begin{center}\mbox{ \epsfxsize 3.0in\epsfbox{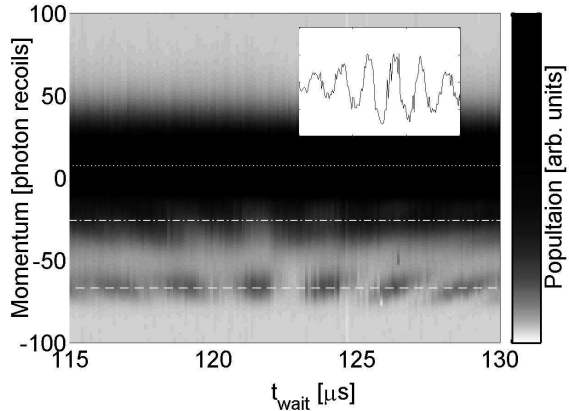}}\end{center}
\caption{Experimental momentum distributions of an ensemble of
cold cesium atoms after two sets of 20 pulses from an off-resonant
standing light wave. Within each set the pulses were separated by
$T=60$ $\mu $s and there was a variable amount of time $t_{{\rm
wait}}$ between the sets. The dotted, dot-dashed and dashed lines
denote atoms accelerated by zero, one and two sets of pulses
respectively. Each vertical slice is the momentum distribution
obtained for one value of $t_{{\rm wait}}$ where the degree of
shading indicates the population. The inset plots the data along
the dashed line (atoms accelerated by both pulses).}
\label{spatial fringes}
\end{figure}

When $t_{\rm wait}$ is scanned in much smaller time steps the
previous experiment can also be used to demonstrate localization
in position. These experiments are only meaningful if they are
performed when the wavefunction has some degree of spatial
localization, that is when $t_{\rm wait}$ is close or equal to an
integer multiple of the pulse separation time. Furthermore, for
these wait times all the elements of the momentum comb described
by Eq.\thinspace (\ref{comb eqn}) translate by an integer number
of standing wave periods (or equivalently acquire a phase of $2\pi
\times \mathop{\rm integer}$). As we have seen, the momentum
distribution produced by the accelerator mode is not a comb of
delta functions. Thus we do not expect an exact rephasing of the
distribution even at the special wait times. Experimentally we
have found that for wait times greater than three pulse separation
times very little rephasing is observed. Figure \ref {spatial
fringes} shows the momentum distributions observed when the pulse
separation time is 60 $\mu $s and the wait time is scanned near to 120 $\mu $%
s. The accelerator mode reappears at the dashed line whenever the
atoms prepared by the first set of pulses have moved a whole
number of standing-wave wavelengths. We can quantify what the
period of the reappearance or revival should be by determining the
distance which the atoms move during $t_{{\rm wait}}$. This
distance is given by $\Delta x=v_{0}t_{{\rm wait}}+ {\frac12}
at_{{\rm wait}}^{2}$, where $v_{0}$\ is the velocity of the atoms
after the end of the first set of pulses. We have calculated this
velocity to be 177 mm/s \cite{revival period}, giving a revival
period of 2.53 $\mu$s. This is in good agreement with the value of
$2.56\pm 0.03$ $\mu$s that is observed along the dashed line in
Fig.\thinspace \ref{spatial fringes}. Another important feature of
this figure is that except for the exact multiple of the pulse
separation time, not all parts of the accelerator mode produced by
the first set of pulses are able to simultaneously undergo
acceleration by the second set of pulses. This can be explained by
noting that the distance moved by an atom during the wait time
depends upon its initial velocity. Thus the revival period for
atoms that have been accelerated the most is the smallest. This
picture also provides an alternative way of understanding why we
only see clear signatures of spatial localization near the
multiples of $T$; the revivals from the different comb elements in
the accelerator mode all come into phase at these times.

The preceding results demonstrate that by using a pulsed periodic
potential it is possible to select restricted regions of phase
space. We now suggest a technique for investigating the dynamical
properties of these regions under the action of a chaotic
potential. Our proposal is to take two atomic states which have
been prepared by an accelerator mode to have identical external
components. We would expose each state to chaotic potentials which
differ slightly in strength and then measure the overlap between
the states in order to determine the dynamics. The $F=3$ and $F=4$
ground state hyperfine levels of cesium offer a convenient means
of realizing such a procedure through a technique analogous to
that of a Ramsey separated field experiment \cite{Ramsey}. After
selecting the desired region of phase space and preparing atoms in
the $F=4$, $m_{{\rm F}}=0$ state, a 50:50 superposition of the two
hyperfine levels can be created by employing a $\frac{\pi }{2}$
pulse of microwave radiation resonant with the 9.19 GHz $F=3$,
$m_{{\rm F}}=0\longrightarrow F=4 $, $m_{{\rm F}}=0$ magnetic
dipole transition. The QDKR light pulses can then be applied using
the same laser beam and pulse envelope as the accelerator mode,
but with the key change of moving the standing wave to make $a=0$
\cite{ResonancePRL}. Importantly, each of the states in the
superposition would experience the standing wave potential with a
different strength. For example, light detuned 30 GHz below
resonance for atoms in the $F=4$ level is detuned approximately 40
GHz for $F=3$ atoms. Since the light shift
scales inversely with detuning, the component of the superposition in the $%
F=3$ state sees a kicking potential which is $\sim 0.75$ the size
experienced by the component in $F=4$. The final step of our
proposal would be to expose the superposition to a second
$\frac{\pi }{2}$ microwave pulse which differs in phase by an
amount $\Delta \phi $\ from the first pulse. If the overlap
between the two final states is good then a
sinusoidal variation in the population of either state would be observed as $%
\Delta \phi $ were systematically changed. The exact value of the
overlap
could be determined by measuring the population at $\Delta \phi =0$\ and $%
\Delta \phi =\frac{\pi }{2}$ and then using a method similar to
that outlined in Ref.\thinspace \cite{Gardiner97}.

To summarize, we have demonstrated a new method of preparing atoms
in a narrow region of phase space from an initially broad
distribution. This was achieved using quantum accelerator modes, a
process in which atoms are exposed to pulses of an off-resonant
standing light wave which is accelerating relative to the atoms'
frame of reference. Since the experimental configuration is very
similar to that used for experiments which have studied the QDKR,
it should be relatively straightforward to apply this technique to
studies of quantum chaos. Previous experiments have concentrated
on the ensemble behavior of the atoms. The work presented here
paves the way for a more detailed examination of the dynamics. We
have also suggested an experiment which would allow the production
of a quantum phase space stability map. By using a method based on
Ramsey separated fields we can find the amount of overlap between
identical initial states which have interacted with chaotic
potentials of slightly different strengths. We hope that the
combination of these two ideas will eventually lead to a fuller
understanding of the concepts of quantum chaos.

\begin{acknowledgments}
We thank A.\thinspace Buchleitner, K.\thinspace Burnett,
S.\thinspace Fishman, S.\thinspace A.\thinspace Gardiner,
I.\thinspace Guarneri, M.\thinspace K.\thinspace Oberthaler and
S.\thinspace Wimberger for stimulating discussions. This work was
supported by the UK EPSRC, the Paul Instrument Fund of The Royal
Society, the EU as part of the TMR `Cold Quantum Gases' network,
contract no.\thinspace HPRN-CT-2000-00125 and the ESF BEC2000+
program.
\end{acknowledgments}


\begin{references}
\bibitem{non-integer am}  The central diffraction order of the accelerator mode will occur at
the nearest integer to $\xi$.
\bibitem{revival period} This velocity is acquired from a free fall
under gravity for 5.2 ms before the beginning of the first set of
accelerator mode pulses, the velocity imparted by the accelerator
mode, and the free fall during the accelerator mode.
\end{references}
\end{document}